\documentclass[conference]{IEEEtran}
\usepackage[utf8]{inputenc}
\usepackage{times,graphics, dsfont, epsfig,amsmath,xspace,endnotes,pifont,multirow,rotating,listings,amssymb,algorithmic,color,caption,nicefrac,adjustbox,tabularx,mathtools, algorithmic,algorithm}
\usepackage{graphicx}
\usepackage{tabularx}

\newcolumntype{L}[1]{>{\raggedright\arraybackslash}p{#1}} 
\newcolumntype{C}[1]{>{\centering\arraybackslash}p{#1}} 
\newcolumntype{R}[1]{>{\raggedleft\arraybackslash}p{#1}} %
\usepackage{steinmetz}
\usepackage{color}
\usepackage[dvipsnames]{xcolor}
\usepackage{amsmath}
\usepackage{amssymb}
\usepackage{float}
\ifCLASSOPTIONcompsoc
    \usepackage[caption=false, font=normalsize, labelfont=sf, textfont=sf]{subfig}
\else
\usepackage[caption=false, font=footnotesize]{subfig}
\fi
\usepackage{graphicx}
\usepackage[letterpaper, left=0.625in, right=0.625in, bottom=1in, top=0.75in]{geometry}

\usepackage{tikz}
\usepackage{amsthm}
\newtheorem{thm}{Theorem}

\newtheorem{problem}[thm]{Problem}

\renewcommand{\H} {{\bf{H}}}
\newcommand{\A} {{\bf{A}}}

\newcommand{\D} {{\bf{D}}}

\renewcommand{\c} {{\bf{c}}}
\newcommand{\g} {{\bf{g}}}

\renewcommand{\a} {{\bf{a}}}
\renewcommand{\r} {{\bf{r}}}
\renewcommand{\d} {{\bf{d}}}

\newcommand{\y} {{\bf{y}}}
\newcommand{\x} {{\bf{x}}}
\newcommand{\z} {{\bf{z}}}

\newcommand{\nariman}[1]{\textcolor{red}{#1}}
\newcommand{\amir}[1]{\textcolor{blue}{#1}}

\author{
\IEEEauthorblockN{Nariman Torkzaban}
\IEEEauthorblockA{\textit{University of Maryland, College Park}\\
College Park, MD\\
narimant@umd.edu}
\and
\IEEEauthorblockN{Mohammad A. (Amir) Khojastepour}
\IEEEauthorblockA{\textit{NEC Laboratories,
America}\\
Princeton, NJ\\
amir@nec-labs.com}
}

\def\BibTeX{{\rm B\kern-.05em{\sc i\kern-.025em b}\kern-.08em
    T\kern-.1667em\lower.7ex\hbox{E}\kern-.125emX}}

\newcommand\copyrighttext{%
  \footnotesize \textcopyright 2021 IEEE. Personal use of this material is permitted.
  Permission from IEEE must be obtained for all other uses, in any current or future
  media, including reprinting/republishing this material for advertising or promotional
  purposes, creating new collective works, for resale or redistribution to servers or
  lists, or reuse of any copyrighted component of this work in other works.
  }
\newcommand\copyrightnotice{%
\begin{tikzpicture}[remember picture,overlay]
\node[anchor=south,yshift=10pt] at (current page.south) {\fbox{\parbox{\dimexpr\textwidth-\fboxsep-\fboxrule\relax}{\copyrighttext}}};
\end{tikzpicture}%
}
\begin{document}

\title{Shaping mmWave Wireless Channel via Multi-Beam Design using Reconfigurable Intelligent Surfaces}




\maketitle
\copyrightnotice
\begin{abstract}
Millimeter-wave (mmWave) communications is considered as a key enabler towards the realization of next-generation wireless networks, due to the abundance of available spectrum at mmWave frequencies. However, mmWave suffers from high free-space path-loss and poor scattering resulting in mostly line-of-sight (LoS) channels which result in a lack of coverage. 
Reconfigurable intelligent surfaces (RIS), as a new paradigm, have the potential to fill the coverage holes by shaping the wireless channel. 
In this paper, we propose a novel approach for designing RIS with elements arranged in a uniform planar array (UPA) structure. In what we refer to as multi-beamforming, We propose and design RIS such that the reflected beam comprises multiple disjoint lobes. Moreover, the beams have optimized gain within the desired angular coverage with fairly sharp edges 
avoiding power leakage to other regions. We provide a closed-form low-complexity solution for the multi-beamforming design. We confirm our theoretical results by numerical analysis.  




\end{abstract}


%
\IEEEpeerreviewmaketitle

\section{Introduction}

Next generation of wireless communication systems aims to address the ever-increasing demand for high throughput, low latency, better quality of service and ubiquitous coverage. The abundance of bandwidth available at the mmWave frequency range, i.e., $[20, 100]$ Ghz, is considered as a key enabler towards the realization of the promises of next generation wireless communication systems. However, communication in mmWave suffers from high path-loss, and poor scattering. Since the channel in mmWave is mostly LoS, i.e., a strong LoS path and very few and much weaker secondary components, the mmWave coverage map includes \emph{blind spots} as a result of shadowing and blockage. Beamforming is primarily used to address the high attenuation in the channel. In addition to beamforming, relaying can potentially be designed to generate constructive superposition and enhance the received signals at the receiving nodes. 
Reconfigurable intelligent surface (RIS)\cite{Huang19}\cite{Liaskos18}\cite{Basar19} is a new paradigm with a great potential for stretching the coverage and enhancing the capacity of next-generation communication systems. Indeed, it is possible to shape the wireless channel by using RIS, e.g., by covering blind spots or providing diversity reception at a receiving node. In particular, passive RIS provide not only an energy-efficient solution but also a cost-effective one both in terms of the initial deployment cost and the operational costs. RIS are promising to be deployed in a wide range of communications scenarios and use-cases, such as high throughput MIMO communications\cite{Huang20}\cite{Nadeem20}, ad-hoc networks, e.g., UAV communications\cite{Li20}, physical layer security\cite{Maka20}, etc. 
Apart from the works focusing on theoretical performance analysis of RIS-enabled systems  \cite{Han19}\cite{Nadeem20}\cite{Jung19}, considerable amount of work has been dedicated to optimizing such an integration, mostly focusing on the phase optimization of RIS elements \cite{Abey20}\cite{Guo20}\cite{Di20}\cite{Ata20} to achieve various goals such as maximum received signal strength, maximum spectral efficiency,  etc. For more information on RIS, we refer the interested readers to \cite{Liu21} and the references therein.



\begin{figure}
    \centering
    \includegraphics[width=0.75\linewidth]{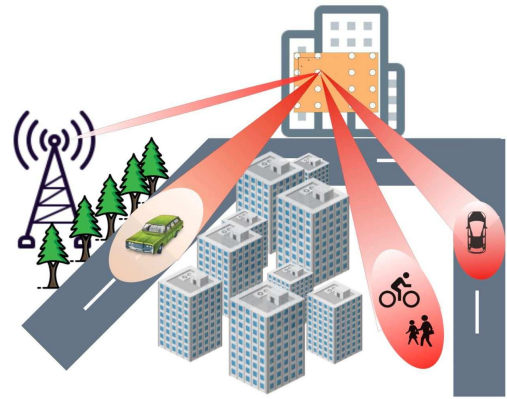}
    \caption{Filling the coverage gap in mmWave communications by utilizing Reconfigurable Intelligent Surfaces enabled by multi-beamforming }
    \label{fig:system}
\end{figure}

In this paper, we consider a communication scenario between a transmitter, e.g., the base station (BS), and terrestrial end-users through a passive RIS that reflects the received signal from the transmitter towards the users. Hence, the users that are otherwise in blind spots of network coverage, become capable of communicating with the base station through the RIS that is serving as a passive reflector (passive relay) maintaining  communication links to the BS and to the users. Given the geo-spatial variance among the locations of the end-users served by the same wireless system, the RIS may have to accommodate users that lie in distant angular intervals simultaneously, with satisfactory quality of service (QoS). In what we refer to as \emph{multi-beamforming}, we particularly address the design of beams consisting of multiple disjoint lobes using RIS in order to cover different blind spots using sharp and effective beam patterns. In the following, we summaries the main contributions of this paper: 
\begin{itemize} 
    \item We design the parameters of the RIS to achieve multiple disjoint beams covering various ranges of solid angle. The designed beams are fairly sharp, have almost uniform gain in the desired angular coverage interval (ACI), and have negligible power transmitted outside the ACI.
    \item We formulate the multi-beamforming design as an optimization problem for which we derive the optimal solution.
    \item Thanks to the derived analytical closed form solutions for the optimal multi-beamforming design, the proposed solution bears very low computational complexity even for RIS with massive array size.
    \item Through numerical evaluation we show that by using passive RIS, multi-beamforming can simultaneously cover multiple ACIs. Moreover, multi-beamforming provides tens of dB power boost w.r.t. single-beam RIS design even when the single beam is designed optimally. 
\end{itemize}

\textbf{Notation} Throughout this paper, $\mathbb{C}$, $\mathbb{R}$, and $\mathbb{Z}$ denote the set of complex, real, and integer numbers, respectively,  $\mathcal{C N}\left(m, \sigma^{2}\right)$ denotes the circularly symmetric complex normal distribution with mean $m$ and variance $\sigma^{2}$, $[a, b]$ is the closed interval between $a$ and $b, \mathbf{1}_{a, b}$ is the $a \times b$ all ones matrix, $\mathbf{I}_{N}$ is the $N \times N$ identity matrix, $\mathds{1}_{[a, b)}$ is the indicator function, $\|\cdot\|$ is the $2$ -norm, $\|\cdot\|_{\infty}$ is the infinity-norm, $|\cdot|$ may denote cardinality if applied to a set and $1$-norm if applied to a vector, $\odot$ is the Hadamard product, $\otimes$ is the Kronecker product, $\mathbf{A}^{H},$ and $\mathbf{A}_{a, b}$ denote conjugate transpose, and $(a, b)^{t h}$ entry of $\A$ respectively.

The remainder of the paper is organized as follows. Section~\ref{sec:desc} describes the system model. In Section~\ref{sec:problem} we formulate the multi-beamforming design problem and propose our solutions in Section ~\ref{sec:proposed}. Section ~\ref{sec:evaluation} presents our evaluation results, and finally, we conclude in Section ~\ref{sec:conclusions}. 



\section{System model} 
\label{sec:desc}
\begin{figure}
    \centering
    \includegraphics[width=0.8\linewidth]{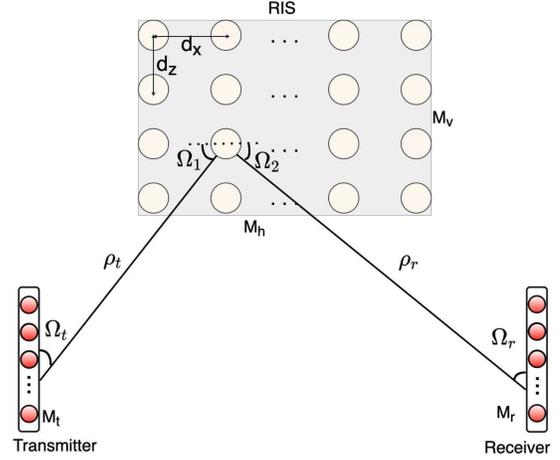}
    \caption{System Model}
    \label{fig:RIS}
\end{figure}

\subsection{Channel Model}
Consider a communication system consisting of a multi-antenna BS with $M_t$ antenna elements as a transmitter and a multi-antenna receiver with $M_r$ antenna elements. The MIMO system is aided by a multi-element RIS consisting of $M$ elements arranged in $M_h \times M_v$ grid in the form of UPA  as shown in figure~\ref{fig:system} where $M_h$ and $M_v$ are the number of elements in the horizontal and vertical directions, respectively. The received signal $\y \in \mathbb{C}^{M_r}$ as a function of the transmitted signal $\x \in \mathbb{C}^{M_t}$ can be written as,
\begin{align}
    \y = (\H_r\boldsymbol\Theta\H_t)\x + \z \label{channel}
\end{align}

\noindent where $\z$ is the noise vector, with each element of $\z$ drawn from a complex Gaussian distribution $\mathcal{C N}\left(0, \sigma_n^{2}\right)$, $\H_t \in \mathbb{C}^{M\times M_t}$ and $\H_r \in \mathbb{C}^{M_r\times M}$ are the channel matrices between each party and the RIS. We assume that the RIS consists of elements for which both the phase $\theta_m$ and the gain $\beta_m$ (in form of attenuation of the reflected signal) of each element, say $m$, may be controlled and $\boldsymbol\Theta \in \mathbb{C}^{M\times M}$ is a diagonal matrix where the element $(m,m)$ denotes the coefficient $\beta_m e^{j \theta_m}$ of the $m^{th}$ element of the RIS. Assuming LoS channel model both between the transmitter and the RIS and between the RIS and the receiver and using the directivity vectors at the transmitter, the RIS, and the receiver, the effective channel matrices can be written as,
\begin{align}
    & \H_r = \a_{M_r}(\Omega_r)\rho_{r}\a^{H}_{M}(\Omega_{2})\label{channel_r} \\
    & \H_t = \a_{M}(\Omega_{1})\rho_{t}\a^{H}_{M_t}(\Omega_{t}) \label{channel_t}
\end{align}

\noindent
where $\a_M(\Omega)$ is the array response vector of an RIS with elements in a UPA structure (RIS-UPA), $\Omega_t$ and $\Omega_2$ are the solid angles of departure (AoD) of the transmitted beams from transmitter and the RIS and $\Omega_1$ and $\Omega_r$ are the solid angle of arrival (AoA) of the received beams at the RIS and the receiver, respectively. 
The gain of the LoS paths from the transmitter to the RIS and from the RIS to the receiver are denoted by $\rho_t$ and $\rho_r$, respectively. Note that the solid angle $\Omega_a$ specifies a pair of elevation and azimuth angles i.e. $(\phi_a, \theta_a)$, $a \in \{1,2, t, r\}$. Further, assuming no pairing between the RIS elements, $\boldsymbol\Theta$ will be a diagonal matrix specified as 
\begin{equation}
    \boldsymbol\Theta = \mbox{diag}\{[\beta_1 e^{j \theta_1}, \ldots, \beta_M e^{j \theta_M}]\}
\end{equation}

\noindent where $\beta_i \in [0,1]$ and $\theta_i \in [0, 2\pi]$.





\subsection{RIS Model}

Suppose an RIS consisting of $M_h \times M_v$ antenna elements forming a UPA structure is placed at the $x$-$z$ plane, where $M = M_h M_v$ and $z$ axis corresponds to horizon. Let $d_z$, and $d_x$ denote the distance between the antennas elements in $z$ and $x$ axis, respectively. 
The directivity of a RIS-UPA can be found in similar way to that of a UPA. At a solid angle $\Omega = (\phi, \theta)$, we have,
\begin{align}
    \a_{M}( \Omega ) = \left[1, e^{j \frac{2 \pi}{\lambda}\r_{\Omega}  \r_1}, \ldots,  e^{j \frac{2 \pi}{\lambda}\r_{\Omega}  \r_{M-1}} \right]^{T} \in \mathbb{C}^{M} \label{directivity_first}
\end{align}
\noindent where respectively,  $\r_{\Omega} = [ \cos \phi \cos\theta, \cos\phi \sin \theta, \sin\phi]$, and $\r_m = (m_hd_x, 0, m_vd_z)$  denote the direction corresponding to the solid angle $\Omega$ and the location of the $m$-th RIS element corresponding to the antenna placed at the position $(m_v, m_h)$. Further, we define a transformation of variables as follows. For a solid angle $\Omega = [\phi, \theta]$, define $\psi = [\xi, \zeta]$ as follows,
\begin{align}
    \xi=\frac{2 \pi d_{z}}{\lambda} \sin \phi \text {,  }\quad \zeta=\frac{2 \pi d_{x}}{\lambda} \sin \theta \cos \phi \label{transformation}
\end{align} Introducing the new variables into equation~\eqref{directivity_first}, it is straightforward to write, 
\begin{align}
    \a_M(\Omega) = \mathbf{d}_{M}\left(\xi, \zeta\right) =
    \mathbf{d}_{M_{v}}\left(\xi\right) \otimes
    \mathbf{d}_{M_{h}}\left(\zeta\right)  \in \mathbb{C}^{M}
\end{align}

where we define for $a \in \{v,h\} $ the  directivity vectors $\d_{M_a}$ as follows, and denote by $\d_M$ the directivity vector corresponding to the RIS. 
\begin{align}
&\mathbf{d}_{M_{v}}\left(\xi\right) = \left[1, e^{j \xi} \cdots e^{j\left(M_{v}-1\right) \xi}\right]^{T} \in \mathbb{C}^{M_{v}}\nonumber\\
&\mathbf{d}_{M_{h}}\left(\zeta\right) = \left[1, e^{j \zeta} \cdots e^{j\left(M_{h}-1\right) \zeta}\right]^{T} \in \mathbb{C}^{M_{h}}\label{directivity_final}
\end{align}



Let $\mathcal{B}$ be the angular range under cover defined as

\begin{equation}
    \mathcal{B} \doteq \left[-\phi^{\mathrm{B}}, \phi^{\mathrm{B}}\right)
    \times \left[-\theta^{\mathrm{B}}, \theta^{\mathrm{B}}\right) \label{range_angle}
\end{equation}

We note that there is a one-to-one correspondence between the solid angle $\Omega = (\phi, \psi)$ and its representation after change of variable as $(\zeta, \xi)$. Accordingly, let $\mathcal{B}^{\psi}$ be the angular range under cover in the $(\zeta, \xi)$ domain given by 

\begin{equation}
    \mathcal{B}^\psi \doteq\left[-\xi^{\mathrm{B}}, \xi^{\mathrm{B}}\right) \times\left[-\zeta^{\mathrm{B}}, \zeta^{\mathrm{B}}\right)
\end{equation}
In this paper, we set $d_x = d_z = \frac{\lambda}{2}$, $\phi^{\mathrm{B}} = \frac{\pi}{4}$, and $\theta^{\mathrm{B}} = \frac{\pi}{2}$, hence $\xi \in [-\pi\frac{\sqrt{2}}{2}, \pi\frac{\sqrt{2}}{2})$, and $\zeta \in [-\pi, \pi)$. Note that, the dependence between variables $\xi$ and $\zeta$ can be resolved using the approximation in \cite{Song17}.
Let us uniformly divide $\mathcal{B}^{\psi}$  into $Q=Q_{v} Q_{h}$ subregions, where $Q_h$ and $Q_v$ are the number of division in horizontal and vertical directions, respectively. A subregion is denoted by  
$$ \mathcal{B}^{\psi}_{ p, q} \doteq \nu_{v}^{p, q} \times \nu_{h}^ {p, q}$$
\noindent where $\nu_{v}^{p} = [\xi^{p-1}, \xi^{p}]$, and $\nu_{h}^{q} = [\zeta^{q-1}, \zeta^{q}]$ defining,
\begin{align}
    & \xi^{p} = -\xi^{\mathrm{B}} + p\delta_v, \quad \zeta^{q} = -\zeta^{\mathrm{B}} + q\delta_h
\end{align}
\noindent with $\delta_v = \frac{2\xi^{\mathrm{B}}}{Q_v}$, and $\delta_h = \frac{2\zeta^{\mathrm{B}}}{Q_h}$. In the next section, we define the multi-beamforming design problem as the core of our proposed RIS structure.




\section{Problem Formulation}
\label{sec:problem}







Prior to formulating the multi-beamforming design problem, we proceed with a few preliminary definitions. Let us define the \emph{multi-beam}  $\mathcal{D} =(\mathcal{D}_1, \ldots \mathcal{D}_k)$ as collection of $k$ \emph{compound beams} $\mathcal{D}_i, i = 1,\ldots, k$ where $\mathcal{D}_i \subseteq \mathcal{B}^{\psi}$ and $\mathcal{D}_i = {\bigcup}_{{(p,q) \in \mathcal{A}_i}} \mathcal{B}^{\psi}_{p,q}$, with $\mathcal{A}_i$ being the set of all pairs $(p,q)$ that all beams $\mathcal{B}^{\psi}_{p,q}$ cover $\mathcal{D}_i$. The union of $\mathcal{B}^{\psi}_{p,q}$ is in fact approximating the shape of the solid angle for the desired compound beam corresponding to $\mathcal{D}_i$. By using larger number of division, i.e., finer beams, one can make the approximation better. We have 
\begin{align}
    &\mathcal{A}_i = \underset{\{\mathcal{\hat{A}}|\mathcal{D}_i \subseteq \underset{(p,q) \in \mathcal{A}}{\bigcup} \mathcal{B}_{p,q}\}}{\arg\min} |\mathcal{\hat{A}}|
\end{align}

Further define $\mathcal{A} = {{\bigcup}_{i=1}^k}\mathcal{A}_i$. 
%
%
We aim to design a beamforming vector $\c$ such that the multi-beam $\mathcal{D}$ is covered when the RIS is excited by an incident wave received at solid angle $\Omega_1$. Using (\ref{channel})-(\ref{channel_t}), the contribution of the RIS in the channel matrix for a receiver at the solid angle $\Omega_2$ is given by
\begin{align}
    \Gamma = \a^{H}_{M}(\Omega_2) \Theta \a_{M}(\Omega_1) = \d^H_M(\Omega_2) \boldsymbol{\lambda}  
\end{align}
where  $\boldsymbol{\lambda} \in \mathbb{C}^M$  is defined as follows. For antenna element located at position $(m_v, m_h)$ in the UPA grid,  we have 
\begin{align}
    \lambda_{m_v, m_h} = \beta_{m_v, m_h}e^{-j(\theta_{m_v, m_h}- m_v\xi_{1} - m_h\zeta_{1})}
\end{align} 
where $(\xi_{1}, \zeta_{1})$ is the representation of $\Omega_1$ in the $\psi$-domain, and hence the vector $\boldsymbol{\lambda}$ is given by
\begin{equation}
    \boldsymbol{\lambda} = [\lambda_{0,0}, \ldots, \lambda_{0,M_h-1}, \lambda_{1,0}, \ldots, \lambda_{M_v-1, M_h-1}]
\end{equation}

We note that $\boldsymbol{\lambda}$ depends on the AoA of the incident beams at the RIS, i.e., $\Omega_1$, as well as the RIS parameters. The reference gain of RIS in direction $(\zeta, \xi) $ in terms of $\boldsymbol{\lambda}$ is given by 
\begin{align}
    G\left(\xi, \zeta, \boldsymbol{\lambda} \right) =\left|\left(\mathbf{d}_{M_{v}}\left(\xi\right)\otimes\mathbf{d}_{M_{h}}\left(\zeta\right)  \right)^{H} \boldsymbol{\lambda} \right|^{2}
\end{align}


On the other hand, the gain of UPA antenna with the feed coefficients $\c$ is given by
\begin{equation}
    G\left(\xi, \zeta, \c \right) =\left|\left(\mathbf{d}_{M_{v}}\left(\xi\right)\otimes\mathbf{d}_{M_{h}}\left(\zeta\right)  \right)^{H} \c \right|^{2} \label{UPA_beamforming_c}
\end{equation}
that has a clear similarity.
This means that to design the RIS-UPA for the STMR problem with receive zone $\mathcal{D}$ we can use the multi-beamforming design framework to cover the ACI's included in $\mathcal{D}$ for the UPA antenna. In particular, a RIS-UPA with parameters $\boldsymbol{\lambda}$ and a UPA-antenna with beamforming parameters $\c$ have the same beamforming gain pattern if UPA structures are the same and $\boldsymbol{\lambda}=\c$. Hence, a RIS-UPA which is excited from the solid angle $\Omega_1$ has the same beamforming gain as its UPA antenna counterpart if $\boldsymbol{\Theta}= \mbox{diag} \{\c^T \odot \a_M^H(\Omega_1)\}$.
For any normalized beamforming vector $\c$, it is straightforward to show that, 
\begin{equation}
    \int_{-\pi}^{\pi} \int_{-\pi}^{\pi} G\left(\xi, \zeta, \mathbf{c}\right) d \xi d \zeta={(2 \pi)^{2}}
\end{equation}

We wish to design beamformers that provide high, sharp, and constant gain within the desired ACI's and zero gain everywhere else. We have then for the ideal gain corresponding to such beamformer $\c$ that,
\begin{align}
&\iint_{\mathcal{B}^{\psi}} G^\text {ideal }_{\mathcal{D}}(\xi, \zeta) d \xi d \zeta =\sum_{i=1}^k\iint_{\mathcal{D}_i} t d \xi d \zeta \nonumber\\
&= \sum_{(p,q) \in \mathcal{A} }{\iint_{\mathcal{B}^\psi_{p,q}} t d \xi d \zeta} = \sum_{(p,q) \in \mathcal{A}}\delta_{p,q} t=(2 \pi)^2 \label{composite}
\end{align}

where $\delta_{p,q}$ denotes the area of the $(p,q)$-th beam in the $(\xi, \zeta)$ domain. Therefore, we can derive $t= \frac{(2\pi)^2}{|\mathcal{A}|\delta_{p,q}}$. It holds that, 
\begin{equation}
    G^{\text {ideal }}_{ \mathcal{D}}\left(\xi, \zeta\right)=\frac{(2\pi)^2}{|\mathcal{A}|\delta_{p,q}} \mathds{1}_{\mathcal{D}}\left(\xi, \zeta\right)\label{ideal_compound}
\end{equation}

Using the beamformer $\c$ we wish to mimic the deal gain in equation \eqref{ideal_compound}. Therefore, we formulate the following optimization problem, 
\begin{align}
& \c^{opt}_{\mathcal{D}} = \underset{\c, \|\c\|=1}{\arg \min } \underset{\mathcal{B}^{\psi}}{\iint}\left|G^{\text {ideal }}_{\mathcal{D}}\left(\xi, \zeta\right)-G\left(\xi, \zeta, \c\right)\right| d \xi d \zeta \label{init_opt}
\end{align}

By partitioning the range of $(\xi, \zeta)$ into the pre-defined intervals, and then uniformly sampling with the rate $(L_v, L_h)$ per interval along both axis,  we can rewrite the optimization problem as follows, 
\begin{align}
 \c^{opt}_{\mathcal{D}} &= \underset{{\c, \|\c\|=1}}{{\arg \min }} \sum_{r=1}^{Q_v}\sum_{s=1}^{Q_h}\iint_{\mathcal{B}^{\psi}_{r,s}}\left|G_{\mathcal{D}}^{\text {ideal }}\left(\xi, \zeta \right)-G\left(\xi, \zeta, \c\right)\right| d \xi d \zeta \nonumber\\&
 = \underset{L_h, L_v \rightarrow \infty}{\lim}\sum_{r=1}^{Q_v}\sum_{s=1}^{Q_h}\sum_{l_v =1}^{L_v}\sum_{l_h =1}^{L_h}\nonumber\\&\frac{\delta_v\delta_h}{L_hL_v}
\left|G_{\mathcal{D}}^{\text {ideal }}\left(\xi_{r, l_v}, \zeta_{s,l_h}\right)-G\left(\xi_{r, l_v}, \zeta_{s,l_h}, \c\right)\right|\label{alternative_opt}
\end{align}
where, 
\begin{align}
    &\xi_{r,l_v} = \xi^{r-1} + l_v\frac{\delta_v}{L_v}, \quad \zeta_{s, l_h} = \zeta^{s-1} + l_h\frac{\delta_h}{L_h} \label{l_v_l_h}
\end{align}

\noindent with $\delta_a = \frac{2\psi_a^{\mathrm{B}}}{Q_a}$, for $a \in \{v, h\}$. Note that it holds for all $(p,q)$ pairs that, $\delta_{p,q} = \delta_v\delta_h$. 
We can rewrite equation \eqref{alternative_opt} as, 
\begin{align}
    \c_{\mathcal{D}}^{opt}=\arg \min _{\c, \|\c\|=1} \underset{L_h, L_v \rightarrow \infty}{\lim}\frac{1}{L_hL_v}\left|\mathbf{G}^{\text {ideal }}_{\mathcal{D}}-\mathbf{G}(\c)\right|\label{init_normed_opt}
\end{align}
where, 
\begin{align}
    \mathbf{G}(\c) =&  \delta_{p,q}\left[G\left(\xi_{1,1}, \zeta_{1,1}, \c\right) \cdots G\left(\xi_{Q_{v}, L_{v}}, \zeta_{Q_{h}, L_{h}}, \c\right)\right]^{T}  
\end{align}
and,
\begin{align}
    \mathbf{G}^{\text {ideal }}_{ \mathcal{D}} =& \delta_{p,q}\left[G^{\text {ideal }}_{ \mathcal{D}}\left(\xi_{1,1}, \zeta_{1,1} \right) \cdots  G^{\text {ideal }}_{ \mathcal{D}}\left(\xi_{Q_{v}, L_v}, \zeta_{ Q_{h}, L_{h}}\right)\right]^{T} 
\end{align}



Unfortunately, the optimization problem in \eqref{init_normed_opt} does not admit an optimal closed-form solution as is, due to the absolute values of the complex numbers existing in the formulation. However, note that, 
\begin{align}
    \mathbf{G}^{\text {ideal }}_{\mathcal{D}}&=\sum_{(p,q) \in \mathcal{A}}\delta_{p,q}\frac{(2\pi)^2}{|\mathcal{A}|\delta_{p,q}}\left(\mathbf{e}_{p,q} \otimes \mathbf{1}_{L, 1}\right) \nonumber\\&= \frac{(2\pi)^2}{|\mathcal{A}|}\sum_{(p,q) \in \mathcal{A}}{\mathbf{e}_{p,q} \otimes \mathbf{1}_{L, 1}}
    \label{ideal}
\end{align}
with $\mathbf{e}_{p,q} \in \mathbb{Z}^{Q}$ being the standard basis vector for the $(p,q)$-th axis among $(Q_v, Q_h)$ pairs. Now, note that $\mathbf{1}_{L, 1}=\mathbf{g} \odot \mathbf{g}^{*}$ for any equal gain $\mathbf{g} \in \mathbb{C}^L$ where $L = L_hL_v$. An equal-gain  vector $\g \in \mathbb{C}^L$ is a vector where all elements have equal absolute values (in this case, equal to $1$). Therefore, we can write: 
\begin{align}
\mathbf{G}^{\text {ideal }}_{\mathcal{D}} &= \sum_{(p,q) \in \mathcal{A}}\frac{(2\pi)^2}{|\mathcal{A}|}\left(\mathbf{e}_{p,q} \otimes\left(\mathbf{g} \odot \mathbf{g}^{*}\right)\right) \nonumber\\
&=\frac{(2\pi)^2}{|\mathcal{A}|}\sum_{(p,q) \in \mathcal{A}}\left(\mathbf{e}_{p,q} \otimes \mathbf{g}\right) \odot\left(\mathbf{e}_{p,q} \otimes \mathbf{g}\right)^{*} \nonumber\\
&=\left(\sum_{(p,q) \in \mathcal{A}}\frac{2\pi}{\sqrt{|\mathcal{A}|}}\left(\mathbf{e}_{p,q} \otimes \mathbf{g}\right)\right)  \nonumber\\&\odot \left(\sum_{(p,q) \in \mathcal{A}}\frac{2\pi}{\sqrt{|\mathcal{A}|}}\left(\mathbf{e}_{p,q} \otimes \mathbf{g}\right)\right)^* \label{final_gik}
\end{align}

Also, it is straightforward to write, 
\begin{align}
    \mathbf{G}(\c)=\left(\D^H \c\right) \odot\left(\D^H \c\right)^{*}\label{dc}
\end{align}

\noindent where, $\D^H = \sqrt{\delta_v\delta_h}(\mathbf{D}_{v}^{H} \otimes \mathbf{D}_{h}^{H})$, and for $a \in \{v,h\}$, and $b= 1\ldots Q_a$ we have, 
\begin{align}
\mathbf{D}_{a} &=\left[\mathbf{D}_{a, 1}, \cdots, \mathbf{D}_{a, Q_{a}}\right] \in \mathbb{C}^{M_{a} \times L_{a} Q_{a}}
\end{align}
where, 
\begin{align}
    &\mathbf{D}_{v, b} =\left[\mathbf{d}_{M_{v}}\left(\xi_{b,1}\right), \cdots, \mathbf{d}_{M_{v}}\left(\xi_{b, L_v}\right)\right] \in \mathbb{C}^{M_{v} \times L_{v}} \\
    &\mathbf{D}_{h, b} =\left[\mathbf{d}_{M_{h}}\left(\zeta_{b,1}\right), \cdots, \mathbf{d}_{M_{h}}\left(\zeta_{b, L_h}\right)\right] \in \mathbb{C}^{M_{h} \times L_{h}}
\end{align}

Comparing the expressions \eqref{init_normed_opt}, \eqref{final_gik}, and \eqref{dc}, one can show that the optimal choice of $\c_\mathcal{D}$ in \eqref{init_opt} is the solution to the following optimization problem for proper choices of $\g_{p,q}$. 

\begin{problem}
Given equal-gain vectors $\g_{p,q} \in \mathbb{C}^L$, for $(p,q) \in \mathcal{A}$  find vector $\c_{\mathcal{D}} \in \mathbb{C}^{M}$ such that
\begin{align}
\c_{\mathcal{D}}=&{\arg \min }_{\c, \|\c\|=1}\nonumber\\& \lim_{L\rightarrow \infty} \left\|\sum_{(p,q) \in \mathcal{A}}\frac{2\pi}{\sqrt{|\mathcal{A}|}}\left(\mathbf{e}_{p,q} \otimes \mathbf{g}_{p,q}\right)- \D^H \c\right\|^{2} \label{obj_func}
\end{align}
\label{main_problem_UPA}
\end{problem}

However, we now need to find the optimal choices of $\g_{p,q}$ that minimize the objective in \eqref{init_normed_opt}. Using \eqref{final_gik}, and \eqref{dc}, we have the following optimization problem.

\begin{problem}
Find equal-gain vectors $\g^*_{p,q} \in \mathbb{C}^L$, $(p,q) \in \mathcal{A}$ such that
\begin{align}
 &<\g^*_{p,q}>_{(p,q)\in\mathcal{A}} = \underset{<\g_{p,q}>_{(p,q)\in\mathcal{A}}}{\arg\min }\nonumber\\
 &\left\| abs(\D^H \c_{\mathcal{D}})- \frac{{2\pi}}{\sqrt{|\mathcal{A}|}} abs(\sum_{(p,q)\in \mathcal{A}}\mathbf{e}_{p,q} \otimes \mathbf{g}_{p,q})\right\|^{2} \label{g_final_eq}
\end{align} 
where $abs(.)$ denotes the element-wise absolute value of a vector.
\label{g_problem}
\end{problem}

Next, we continue with the solution of Problems ~\ref{main_problem_UPA}, and~\ref{g_problem}.

\section{Proposed Multi-beam Design}
\label{sec:proposed}

Note that the solution to problem \ref{main_problem_UPA} is the limit of the sequence of solutions to a least-square optimization problem as $L$ goes to infinity. For each $L$ we find that,
 \begin{align}
& {\c}^{(L)}_{\mathcal{D}} = \sum_{(p,q) \in \mathcal{A}}\frac{2\pi}{\sqrt{|\mathcal{A}|}}(\D \D^H)^{-1} \D  \left(\mathbf{e}_{p,q} \otimes \mathbf{g}_{p,q}\right) \\
& {\c}^{(L)}_{\mathcal{D}} = \sum_{(p,q) \in \mathcal{A}}\sigma \D_{p,q}\g_{p,q} \label{c_final_eq}
\end{align}
where $\sigma = \frac{{ 2\pi \sqrt{\delta_{v}\delta_{h}} }}{L Q\delta_{v}\delta_{h}\sqrt{|\mathcal{A}|}} = \frac{2\pi}{LQ\sqrt{\delta_{v}\delta_{h}|\mathcal{A}|}}$, noting that it holds that, 
\begin{align}
\D\D^H &= \delta_v\delta_h(\D_v \otimes\D_h)(\D_v^H \otimes \D_h^H) = \delta_v\delta_hLQ
\end{align}


Even though Problem~\ref{main_problem_UPA} admits a nice analytical closed form solution, doing so for the Problem~\ref{g_problem} is not a trivial task, especially due to the fact that the objective function is not convex. However, the convexification of the objective problem \eqref{g_final_eq} in the form of 
\begin{align}
 \underset{<\g_{p,q}>_{(p,q)\in\mathcal{A}}}{\arg\min }
 \left\| \D^H \c_{\mathcal{D}}- \frac{{2\pi}}{\sqrt{|\mathcal{A}|}} \sum_{(p,q)\in \mathcal{A}}\mathbf{e}_{p,q} \otimes \mathbf{g}_{p,q} \right\|^{2} \label{g_final_eq_convexified}
\end{align} 
and using $\c_{\mathcal{D}}$ from \eqref{c_final_eq} leads to an effective solution for the original problem.
Indeed, it can be verified by solving the optimization problem \eqref{g_final_eq_convexified} numerically that a close-to-optimal solution admits the following form.
\begin{align}
    \g^*_{p,q} = \left[{ 1,  \alpha_v \alpha_h, \cdots,  \alpha_v^{ (L_v -1)}\alpha_h^{ (L_h -1)} }\right]^T, (p,q) \in \mathcal{A} \label{g_conjecture}
\end{align}
for some $\eta_v$, $\eta_h$  where $\alpha_a = e^{j(\frac{\eta_a}{L_a})}$, $a \in \{v, h\}$. \label{proposiiton_g}
%
In the following, we use the the analytical form \eqref{g_conjecture} for $\g^*_{p,q}$ for the rest of our derivations. This solution would not be the optimal solution for the original problem \eqref{g_final_eq}.  However, it provides a near optimal solution with added benefits of allowing to (i) find the limit of the solution as $L$ goes to infinity, and (ii) express the beamforming vectors in closed form, as it will be revealed in the following discussion. An analytical closed form solution for $\c_\mathcal{D}$ can be found as follows. It holds that, 
\begin{align}
     &{\c_{\mathcal{D}}}^{(L)}  =  \sum_{(p,q)\in \mathcal{A}}\left(\sum_{(l_v, l_h)=(1,1)}^{(L_v, L_h)}\sigma g_{p,q, l_v, l_h}\mathbf{d}_{M_t}\left(\xi_{p,l_v}, \zeta_{q,l_h}\right) \right) \nonumber\\
     & = \sum_{(p,q)\in \mathcal{A}}\left(\sum_{(l_v, l_h)=(1,1)}^{(L_v, L_h)}  \sigma g_{q,p,l_v, l_h}\left[1, \cdots,  e^{j\mu_{p,q, l_v, l_h}^{M_v-1, M_h-1}}\right]^T  \right) 
\end{align}

\noindent where $ \mu_{p,q, l_v, l_h}^{m_v, m_h} = \left( m_v\xi_{p,l_v} + m_h\zeta_{q,l_h}\right)$. We can then write for the $(m_v, m_h)^{th}$ component of the beamformer $\c_{\mathcal{D}}$, 
\begin{align}
    &c_{p,q, m_v, m_h} \nonumber\\&=  \lim_{L_h, L_v\rightarrow \infty} \frac{1}{L_hL_v}\sum_{(p,q)\in \mathcal{A}}\sum_{(l_h, l_v)=(1,1)}^{(L_h, L_v)} g_{p,q, l_v, l_h}e^{j\mu_{p,q, l_v, l_h}^{M_v-1, M_h-1}} \label{c_middle}
\end{align}

Using equation \eqref{l_v_l_h}, we can rewrite \eqref{c_middle} as,
\begin{align}
    c_{p,q, m_v, m_h} =&  \frac{2\pi}{Q}e^{j\chi_{p-1, q-1}^{m_v, m_h}}
    \left(\frac{1}{L_v}\lim_{ L_v\rightarrow \infty} \sum_{l_v=1}^{L_v} e^{j\frac{\eta_v+ m_v\delta_v}{L_v} l_v}\right) \nonumber\\&
    \left(\frac{1}{L_h}\lim_{ L_h\rightarrow \infty} \sum_{l_h=1}^{L_h} e^{j\frac{\eta_h+ m_h\delta_h}{L_h} l_h}\right)
\end{align}

to get, 
\begin{align}
    c_{\mathcal{D}, m_v, m_h} &=  \sum_{(p,q) \in \mathcal{A}}\frac{2\pi}{Q}e^{j\chi_{p-1, q-1}^{m_v, m_h}}
    \int_{0}^{1} e^{j\xi_v x}dx\int_{0}^{1} e^{j\xi_h x}dx \nonumber\\&
    = \sum_{(p,q) \in \mathcal{A}}\frac{2\pi}{Q}e^{j(\zeta_{p-1, q-1}^{m_v, m_h}+ \frac{\xi_v+\xi_h}{2})} sinc(\frac{\xi_v}{2\pi})sinc(\frac{\xi_h}{2\pi})
\end{align}

\noindent with $ \chi_{p,q}^{m_v, m_h} = \left( m_v\xi^{p} + m_h\zeta^{q}\right)$, and $\xi_a = {\delta_a}{m_a} + \eta_a$,  for $a\in \{v,h\}$. Now that the closed-form expression for $\c_{\mathcal{D}}$, and therefore, $\boldsymbol{\lambda}$  is known, the RIS parameters at the antenna placed at location $(m_v, m_h)$ can be easily computed. More precisely, we get, 
\begin{align}
    &\beta_{m_v, m_h} = {|\c_{\mathcal{D}, m_v, m_h}|}\\
    & \theta_{m_v, m_h} = \phase{\c_{\mathcal{D}, m_v, m_h}} + m_v\psi_{v,1} + m_h\psi_{h,1}
\end{align}

In the case that gain control (attenuation) at the RIS elements is not feasible, $\beta_{m_v, m_h} = 1$ will be replaced by the derivation for the absolute value of the RIS parameters. Next, we verify the effectiveness of our multi-beamforming design approach by means of numerical experiments. 


\section{Performance Evaluation}
\label{sec:evaluation}

\begin{figure*} 
    \centering
        \subfloat[Dual-beam 3D UPA Pattern \label{one-sided64}]{
            \includegraphics[width=0.26\linewidth]{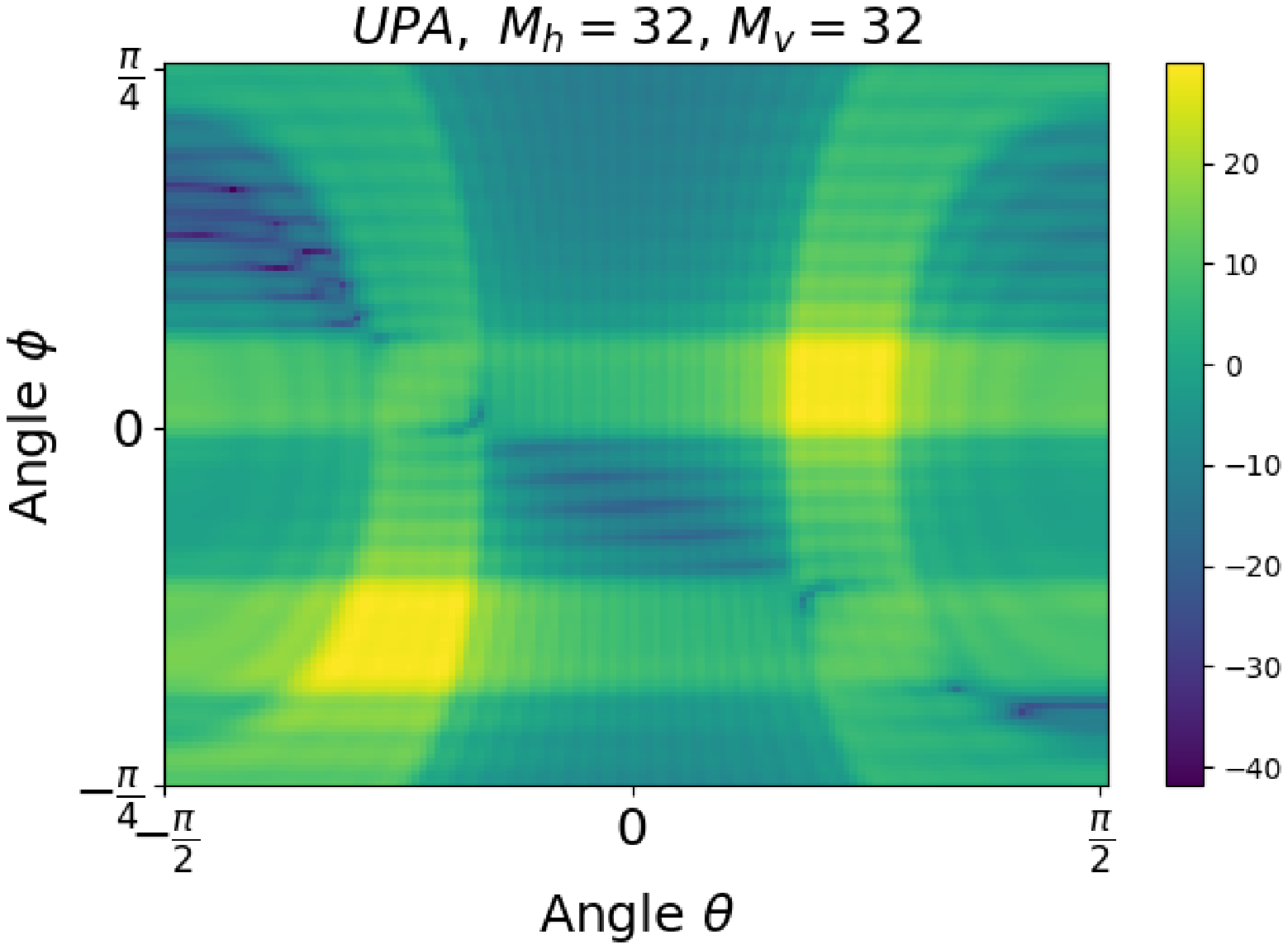}}
        \subfloat[UPA Pattern Cut at $\phi_c$ \label{TULA_H_64}]{
            \includegraphics[width=0.25\linewidth]{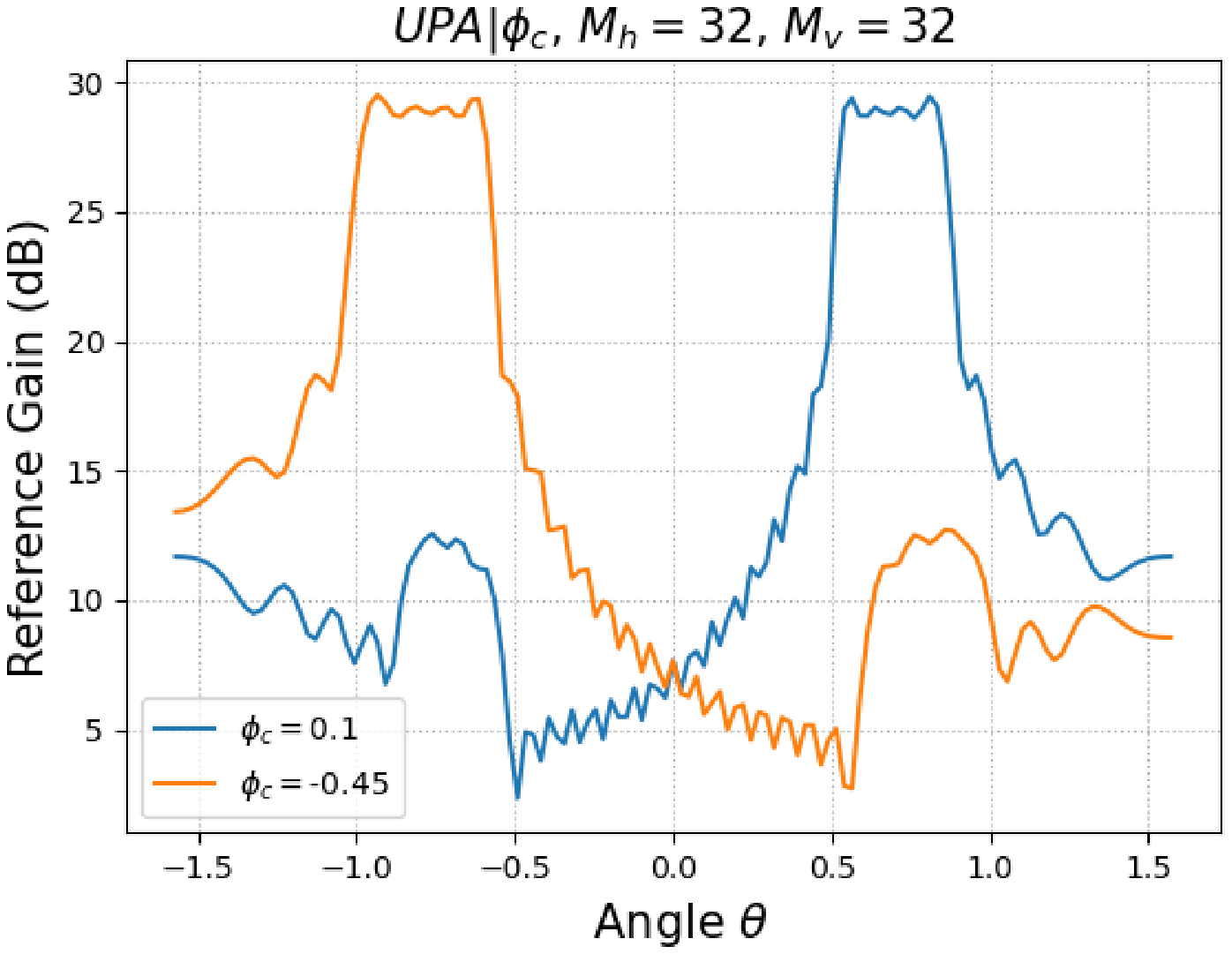}}
        \subfloat[UPA Pattern Cut at $\theta_c$ \label{fig:1024omp_comp}]{
        \includegraphics[width=0.25\linewidth]{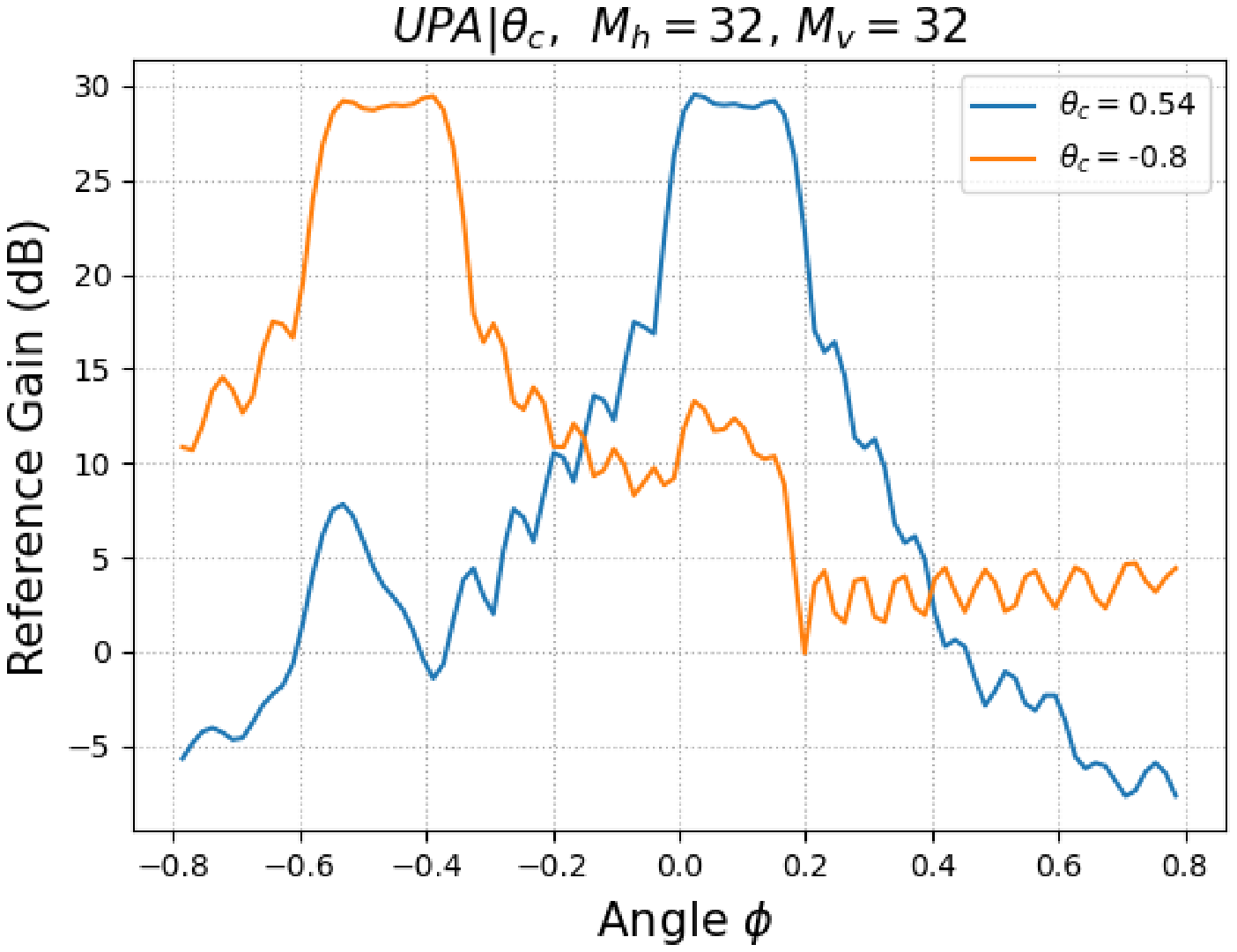}}
        \subfloat[3D UPA Pattern \label{one-sided32}]{
            \includegraphics[width=0.26\linewidth]{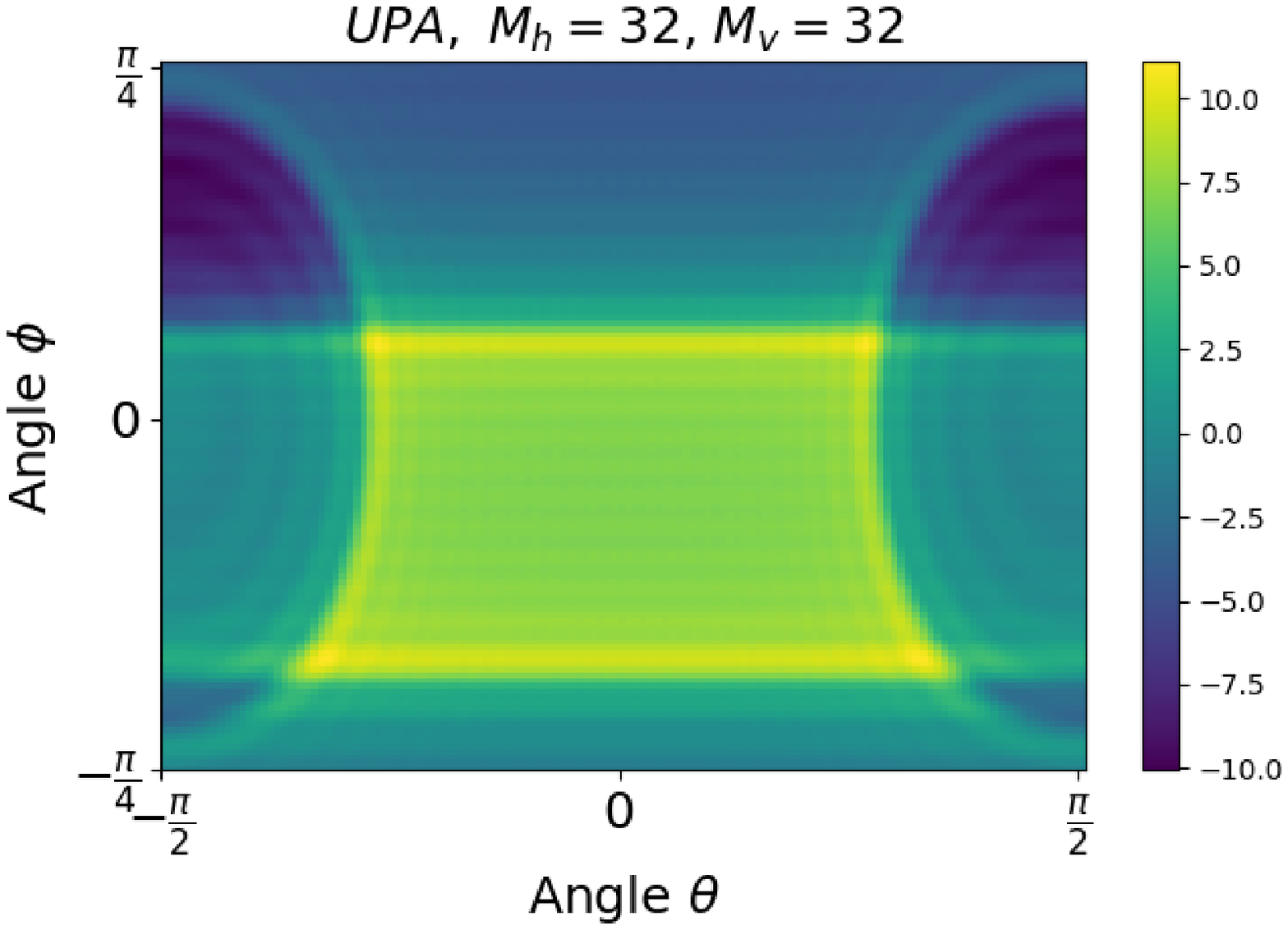}}
  \caption{Beam pattern for UPA structure}
  \label{fig:UPA_pattern} 
\end{figure*}


In this section, we evaluate the performance of our multi-beam design framework. We aim to design a dual-beam  which comprises of two lobes with centers at $(-8\pi/32, -5 \pi /32)$ and $(7\pi/32, \pi /32)$ in the $(\phi, \theta)$ domain, and with the beam-width equal to $\pi /16$. We divide both the $\psi_h$, and the $\psi_v$ range uniformly into $Q_h = 16$, and $Q_v = 16$ regions resulting in $Q = 256$ equally-shaped units in $(\psi_v, \psi_h)$ domain. We cover each desired beam with the smallest number of the designed units to provide uniform gain at the desired angular regions.
Figures~\ref{fig:UPA_pattern}(a)-(c) depict the beam pattern of the dual beam obtained through our design where all angles are measured in radians. 
Figure~\ref{fig:UPA_pattern}(a), shows the heat-map corresponding to the gain of the reflected beam from RIS for the designed dual-beam. The gains are computed in dB. It can be seen that the designed beamformer generates two disjoint beams with an almost uniform gain over the desired ACIs. It is also observed that the beams sharply drop outside the desired ACIs and effectively suppress the gain everywhere outside ACI. In order to quantify the suppression we depict the cross-section of the gain pattern at a fixed elevation angle $\phi_c$ for two values of $\phi_c \in \{-8\pi/32, 7\pi/32\}$ located inside the two lobes of the designed dual beam in Figure~\ref{fig:UPA_pattern}(b). Similarly, Figure~\ref{fig:UPA_pattern}(c) shows the cross-section of the beam pattern at a fixed azimuth angle $\theta_c$ for two values of $\theta_c \in \{ -5 \pi /32, \pi /32\}$. Both Figures~\ref{fig:UPA_pattern}(b)~and~(c) confirm the sharpness of both lobes of the designed dual-beam and can be used to find the beamwidth of each lobes at an arbitrary fraction from its maximum values, e.g,, the 3dB beamwidth or 10dB beamwidth. Indeed, there is negligible difference between 3dB and 10dB beamwidth which clarifies the sharpness of the beams. From Figures~\ref{fig:UPA_pattern}(b)~and(c), it is also observed that the gain within the ACI is almost uniform. Nonetheless, we should emphasize the fact that the shape of the lobes of the beam that are centered at different solid angle may suffer from slight deformation as seen by Figure~\ref{fig:UPA_pattern}(a). This phenomenon worsens as the corresponding lobes of the beams get too close to the plane of the RIS. 

Finally, in order to compare the performance of our multi-beam design to a single-beam design, we consider a beam with single lobe which is capable of covering the same two regions as in the dual-beam design. Figure~\ref{fig:UPA_pattern}(d), shows the heat-map corresponding to the gain of the reflected beam from RIS for the corresponding single beam that is optimized based on our design. As it was the case for multi-beam, this figure also shows that for a single beam our design generates an almost uniform and fairly sharp beam. However, comparing Figures~\ref{fig:UPA_pattern}(a)~and(d), we observe that in the desired ACI, the multi-beamforming procedure, enhances the gain by about $20$ dB over the beams with optimized single lobe.

\section{Conclusions}
\label{sec:conclusions}
RIS can be incorporated into mmWave communications to fill the coverage gaps in the blind-spots of the mmWave system. We proposed a novel approach for designing RIS employing a UPA antenna structure, that is capable of covering multiple disjoint angular intervals simultaneously. Both our theoretical results and numerical experiments demonstrate that our technique termed as multi-beamforming will result in sharp, high, and stable gains within the desired ACI's regardless of their spatial locations, while effectively, suppressing all the undesired out-of-band components.



\renewcommand{\nariman}[1]{\textcolor{red}{#1}}
\renewcommand{\amir}[1]{\textcolor{blue}{#1}}



%

\bibliographystyle{IEEEtran}
\bibliography{bibliography}

\end{document}